\documentclass{WileyMSP-template}
\usepackage{moreverb,url}
\usepackage{multicol}
\usepackage{tabularx}
\usepackage[normalem]{ulem} % for strikethrough text
\usepackage{multicol}
\usepackage{multirow}
\usepackage{subfigure}
\usepackage{booktabs}
\usepackage{xtab}
\usepackage{scrextend}
\usepackage{rotating}
\usepackage[symbol]{footmisc}
\def\@fnsymbol#1{\ensuremath{\ifcase#1\or *\or \dagger\or \ddagger\or
   \mathsection\or \mathparagraph\or \|\or **\or \dagger\dagger
   \or \ddagger\ddagger \else\@ctrerr\fi}}
\usepackage{graphicx}  % <----
\usepackage{hyperref}
\hypersetup{colorlinks,bookmarksopen,bookmarksnumbered,citecolor=black,urlcolor=black}
\usepackage{mathtools}
\usepackage{makecell}
\usepackage[style=apa, backend=biber]{biblatex}
\usepackage[T1]{fontenc}
\begin{document}

%\pagestyle{fancy}
%\rhead{\includegraphics[width=2.5cm]{vch-logo.png}}

\title{\noindent TRACE-Omicron: Policy Counterfactuals to Inform Mitigation of COVID-19 Spread in the United States}

\maketitle

% Author: Please give full first and last names for authors and include * after the name of all corresponding authors
\author{David O'Gara\footnote{These authors contributed equally to this work\label{authorfootnote}}}
\author{Samuel F. Rosenblatt\footref{authorfootnote}}
\author{Laurent Hébert-Dufresne}
\author{Rob Purcell}
\author{Matt Kasman}
\author{Ross A. Hammond\footnote[2]{
To whom correspondence should be addressed.}}

% Affiliations: Please provide adacemic titles (Prof. or Dr.) for all authors where applicable, and include an institutional email address for all corresponding authors
\begin{affiliations}
\noindent Center On Social Dynamics and Policy, Brookings Institution (R.A.H, M.K, R.P) \\
\noindent Division of Computational and Data Sciences, Washington University in St. Louis (D.O.G, R.A.H) \\
\noindent Vermont Complex Systems Center, University of Vermont (S.F.R, L.H.-D.) \\
\noindent Department of Computer Science, University of Vermont (S.F.R, L.H.-D.) \\
\noindent Brown School, Washington University in St. Louis (R.A.H) \\
\noindent Santa Fe Institute (R.A.H)\\
\noindent Email Address:
\noindent rhammond@brookings.edu

\end{affiliations}

% Keywords: Please provide a minimum of three and a maximum of seven keywords, separated by commas

%\keywords{Keyword 1, Keyword 2, Keyword 3}

% Abstract should be written in the present tense and impersonal style (i.e., avoid we), and be at most 200 words long
\begin{abstract}

The Omicron wave was the largest wave of COVID-19 pandemic to date, more than doubling any other in terms of cases and hospitalizations in the United States. In this paper, we present a large-scale agent-based model of policy interventions that could have been implemented to mitigate the Omicron wave. Our model takes into account the behaviors of individuals and their interactions with one another within a nationally representative population, as well as the efficacy of various interventions such as social distancing, mask wearing, testing, tracing, and vaccination. We use the model to simulate the impact of different policy scenarios and evaluate their potential effectiveness in controlling the spread of the virus. Our results suggest the Omicron wave could have been substantially curtailed via a combination of interventions comparable in effectiveness to extreme and unpopular singular measures such as widespread closure of schools and workplaces, and highlight the importance of early and decisive action.

\end{abstract}

% Text: Please use section headings and subheadings as specified below. For communications, all section headings apart from Experimental Section should be removed
% Please make the first reference to a display item bold: \textbf{Figure 1}
% Do not abbreviate Figure, Equation, etc.; display items are always singular, i.e., Figure 1 and 2.
% Equations are always singular, i.e., Equation 1 and 2, and should be inserted using the {equation} environment, not as graphics
% Please do not use footnotes in the text, additional information can be added to the Reference list.

\section{Introduction}

The COVID-19 pandemic continues to pose a threat to the health and stability of our society. From December 29, 2021 to February 27, 2022, the Omicron variants of COVID-19 caused over 100,000 deaths and over 30 million new reported infections \parencite[]{cdc_covid_data_tracker}. The negative impact of the ``Omicron wave'' has lasted beyond the winter of 2021-2022, including a high and growing burden of chronic disease from long-COVID \parencite{bach_new_2022}, as well as continued social and economic disruption \parencite{glennerster_calculating_2022,bureau_of_labor_statistics_effects_nodate}.

Disease prevention strategies have evolved since the emergence of COVID-19 in early 2020. The population now has access to a wide selection of mitigation tools: vaccines which are effective at preventing death, a larger stock of N95 masks, and multiple forms of diagnostic testing. However, the SARS-CoV-2 virus continues to evolve, both in transmissibility as well as immune evasion. New variants are already beginning to emerge, prompting predictions of a potential winter/spring wave \parencite{callaway_will_2022}. We can prepare for the next phase—and those that are likely to follow—by using last winter’s Omicron wave to help understand what might happen and how we can best respond to mitigate harm. 

We have developed an agent-based model that embraces the complexity of current outbreaks and mitigation strategies, as well as the uncertainty about the future course the pandemic will take. Classic compartmental models are powerful because of their simplicity, but fail to capture strategic trade-offs that emerge in real scenarios. For example, when simulating interventions in compartmental models, both the scale of an epidemic peak (maximum number of cases at a given time) and the size of an epidemic (total number of cases over time) will decrease monotonically together. Because reality is more complex, an intervention might improve one while worsening the other. Similarly, these compartmental models also fail to capture the heterogeneity of real contact networks and therefore assume that all interventions will impact all individuals in similar ways. Again, reality is more complex, and our models should therefore account for contact patterns.

In heterogeneous, networked, populations, we want to model different policies aimed at mitigating the number of cases over time. These policies can use an array of behavioral, technological, and biomedical strategies; most importantly social distancing, masking, and vaccination. We focus on cases and not hospitalizations or deaths for a few reasons. One, we will show that it is sufficient to capture rich strategic trade-offs in policy design. Two, risk of complications could be considered as a fraction of cases and will therefore be approximately proportional to number of cases. Three, other long-terms consequences of COVID-19 are not yet fully understood and focusing on certain complications could limit the applicability of the model. 

 Agent-based modeling has a long history in pandemic response planning in the United States \parencite{germann_mitigation_2006,epstein_modelling_2009}, and the COVID-19 pandemic is no exception \parencite{kerr_covasim_2021,bedson_review_2021}. To glean actionable insights from the Omicron wave, we extend and apply TRACE, a nationally representative agent-based simulation \parencite{hammond_modeling_2021} which models COVID-19 dynamics in highly realistic contact settings. Like other agent-based simulations, our model, TRACE-Omicron, allows disaggregated modeling of individual agent interactions. This allows us to incorporate large amounts of heterogeneity in disease transmissibility, contact structure, and policy interventions. 

Our key contribution to the COVID-19 pandemic modeling literature is applying a sophisticated model to the interaction of a wide array of possible disease scenarios and potential response strategies—here, including strategies which were not used, but could have been. By looking backward at the Omicron wave, we can consider how different policy combinations might have changed the observed outcome. Because we also vary disease conditions, we can identify response strategies that might be best suited to specific future variations that are yet to be observed or are robust across disease conditions and thus useful to employ in the face of epidemiological uncertainty. 

Our research suggests feasible strategies that can effectively limit infection rates across a wide range of potential COVID-19 variants. For example, increased use of high quality masks or combining several policies at a lower intensity, such as testing, mask efficacy, and boosting, can outperform singular high intensity interventions.  Our results also allow consideration of potential policy substitutes, allowing decision-makers to quantitatively assess potential alternatives to disruptive or impractical options.  In this paper, we highlight these strategies in the hope that policymakers will use them as part of a ``response playbook'' that will aid in rapid, timely, and effective action that can limit damage as COVID-19 progresses. 

\section{Methods}

\subsection{Model Design and Dynamics}

TRACE-Omicron builds upon prior models \parencite{hammond_modeling_2021} to consider what potential impact different policy interventions might have had upon the (BA.1, lineage) Omicron wave. TRACE-Omicron tracks individual agents: simulated individuals existing in realistic social structures whose actions shape, and are shaped by, their interactions with other agents. An agent’s ``state'' includes their current disease, quarantine and testing statuses, their vaccination history, whether they are wearing a mask, and who they interact with. Opportunities for disease transmission are driven by agent states and interactions, and are shaped by policy interventions. An agent who tests positive for COVID-19, or is in contact with someone who tests positive, is expected to quarantine and not interact with other agents for 10 days, or 5 if they are asymptomatic \parencite{centers_for_disease_control_and_prevention_covid-19_quarantine}. We allow for some amount of non-adherence, reflecting the fact that some individuals may be unable or unwilling to comply with quarantine procedures. Other interventions, such as vaccination or masking, reduce the probability of infection for a non-infected agent, and also reduce the probability of onward transmission for an agent who experienced an infection. We simulate many combinations of these interventions to understand how these policies interact with one another, and which ones, at which intensities, are particularly effective at reducing disease spread. In total, we conducted 88,128,000 model runs exploring 46,080 parameter combinations. We describe the functionality of TRACE-Omicron in the following sections. A full description of TRACE-Omicron is given in the Model Sketch, found in the Supporting Information. 

\subsection{Disease Progression}

\begin{figure}
    \centering
    \includegraphics{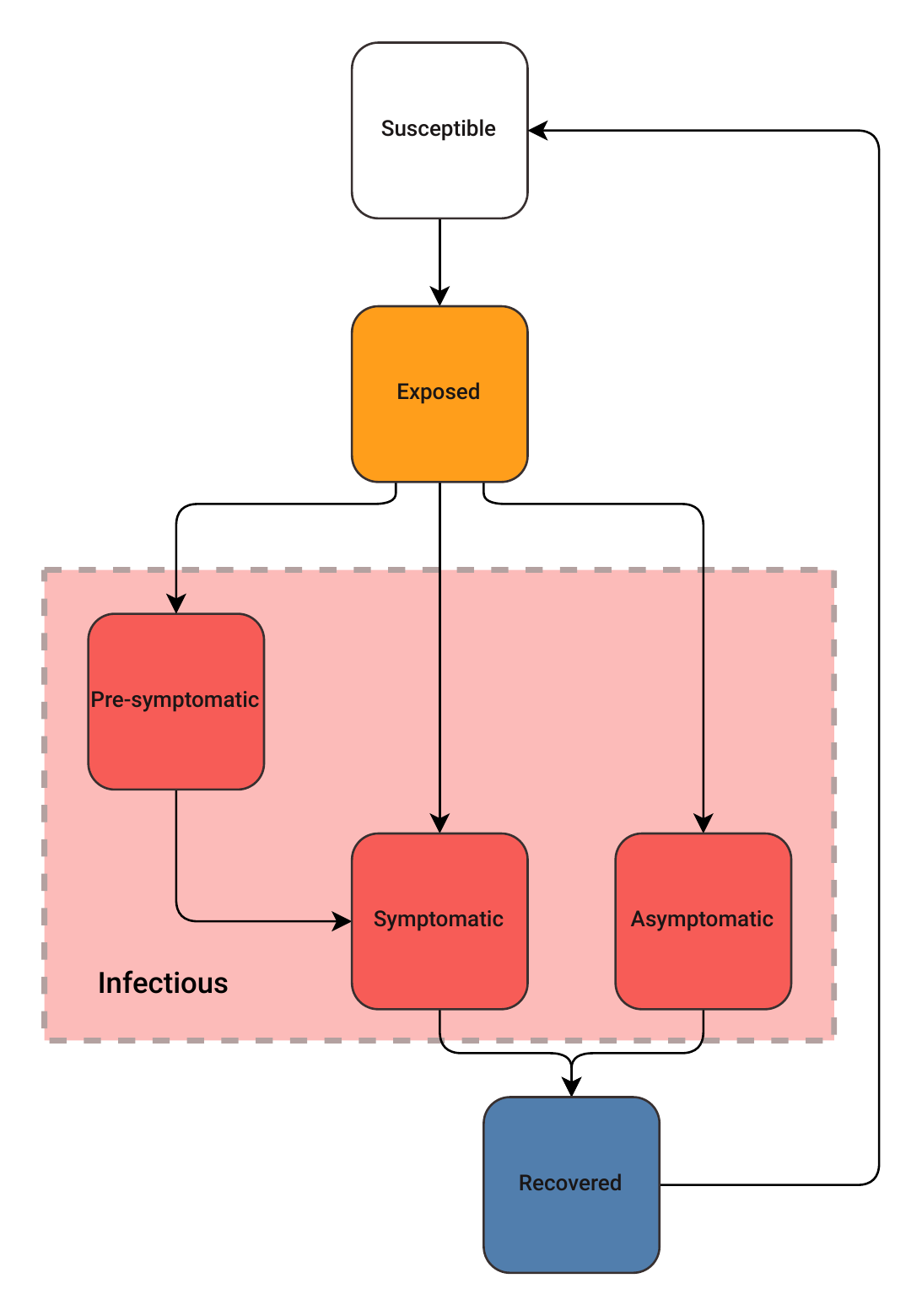}
    \caption{Flow chart of COVID-19 ``states'' and possible ``state transitions'' represented in the model.}
    \label{fig:compartments}
\end{figure}

Infection progression in our simulations uses a variant of the classic ``Susceptible-Exposed-Infectious-Recovered'' (SEIR) epidemiological model that is intended to specifically represent COVID-19, see Fig.~\ref{fig:compartments}. Agents may be in one of several disease states and progress through each state via interactions with other agents. Agents who have never experienced a COVID-19 infection or no longer have substantial antibody protection from a prior infection are ``susceptible'' and move to ``exposed'' after contact with an infectious agent. After a set incubation period, they become ``infectious,'' with some individuals having a shorter incubation and are infectious before they show symptoms (i.e., are ``pre-symptomatic'') and others never show symptoms and are called asymptomatic \parencite{song_serial_2022,ge_covid-19_2021,du_serial_2020,kimball_asymptomatic_2020,wei_presymptomatic_2020,laitman_sars-cov-2_2022,allen_comparative_2022,ferguson_report_2020}. Infectious agents progress through their disease state for a set time until they are ``recovered.'' After a period of immunity, agents can probabilistically transition back to being susceptible based on their time since infection \parencite{townsend_durability_2021}. To capture a critical feature of COVID-19, we allow for infectivity to vary across individuals and infection type, allowing for both highly contagious ``super-spreaders'' as well as a lower likelihood of non-symptomatic individuals transmitting the disease. The number of agents in each disease state and their time in each state is based on CDC data from late December 2021, adjusted for undercounting \parencite{centers_for_disease_control_and_prevention_estimated_burden}.

\subsection{Calibration and Initialization}

\begin{figure}
    \centering
    \includegraphics[width=\linewidth, trim={0 0 2cm 0},clip]{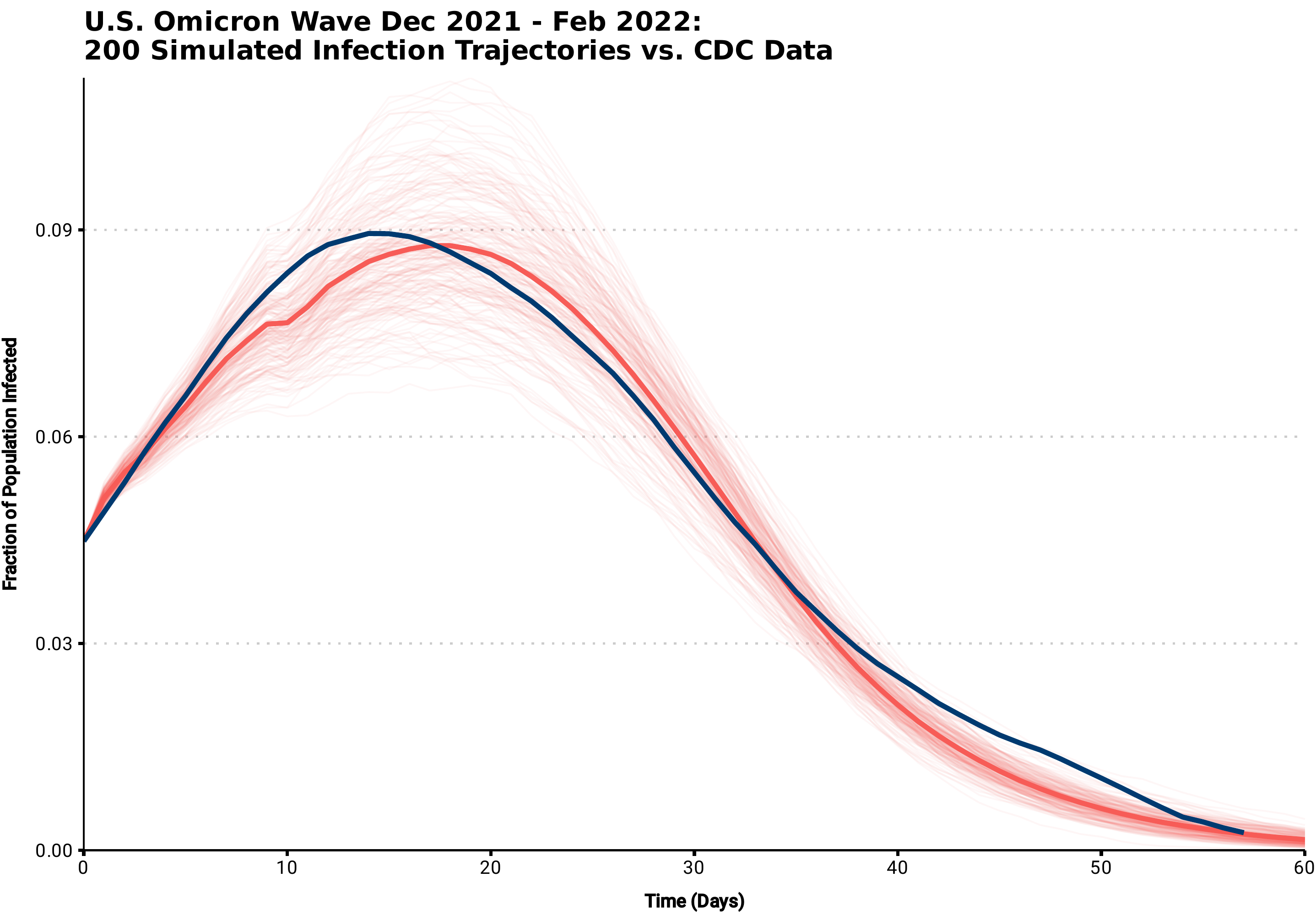}
    \caption{Estimation of baseline conditions under Best Fit calibration. Each curve represents the count of infected agents each tick, compared to estimates using CDC data. Individual trace lines represent 200 simulations, with the mean infected agents per day highlighted in bold.}
    \label{fig:calibration}
\end{figure}

TRACE-Omicron is initialized to reflect the state of COVID-19 spread in the United States in late December 2021. During the Omicron wave, reinfections with SARS-CoV-2 were common. We model three key features of the Omicron wave based on available literature: the amount of prior immunity in the population, the waning of antibody protection over time, and imperfect protection conferred by prior infection against reinfection by Omicron. We use a ``burn-in'' procedure to start a model run during a pandemic, rather than being forced to study an initial outbreak of infections. Using CDC case and vaccination data, we estimate the number of infected agents on December 29, 2021, how long each has been in their initial state, and the actions they would have taken in the days leading up to the start of the model run. We also initialize the baseline intensities of policy interventions in the United States in place in late December 2021.

A standard test for ABMs is whether they can achieve ``generative sufficiency,'' that is, whether an ABM can explain empirical phenomena \parencite{joshua_m_epstein_generative_2006}. To calibrate TRACE-Omicron, we ground all parameters empirically as far as possible, but allow to vary those parameters which are difficult to measure directly, show substantial variance in estimates across multiple studies, or require adjustment for compatibility with the model. Specifically, we vary the proportion of asymptomatic cases, the false negative rate of antigen tests, daily contact tracing capacity \parencite{simmons-duffin_covid-19_2020,simmons-duffin_covid-19_2020-1}, the rate of mask wearing in the population (see Model Sketch), the base transmission probability of infection \parencite{reuters_fact_2022}, and the immune evasion probability \parencite{carazo_estimated_2022,altarawneh_protection_2022,ferguson_report_2021}.

To test our model robustness, we explore three calibration scenarios in our simulations, summarized below and in Table~\ref{tab:baselines}:
\begin{itemize}
    \item The ``Best Fit'' calibration, which has the lowest mean-squared error when compared to the CDC data.
    \item The ``Tractable Strain'' calibration was chosen as an alternative that also had substantial literature and empirical support, and was representative of a class of parameterizations with lower base transmission rate and lower antigen test false negative rate. As a result, lower levels of policy intervention are required at baseline for similar calibration outcomes.    
    \item The ``High Immune Escape'' calibration, which has the lowest mean-squared error among high immune escape scenarios. We explore this as a sensitivity test, due to ongoing uncertainty about the immune evasion of the variants in the Omicron family.
\end{itemize}

For each calibration, we calculate the mean-squared error between the number of infected agents per model step and CDC data. To establish generative sufficiency, we track the number of infections during the Omicron wave, as shown in Fig.~\ref{fig:calibration}. In subsequent sections, we primarily present the results of the Best Fit calibration, but we also explore the other calibrations as a robustness check throughout.

\begin{table}
\centering
\begin{tabular}{lllll}
\toprule
\textbf{Parameter} &  \textbf{\makecell{Best \\ Fit}} & \textbf{\makecell{Tractable \\ Strain}} & \textbf{\makecell{High Immune \\ Escape}} & \textbf{\makecell{Calibration  \\ Search Space}} \\ \hline
Antigen False Negative Rate &           0.25 &     0.2 &               0.2 & {0.2, 0.25, 0.3} \\
       Daily Trace Capacity &         250 &   250 &                250  & {50, 250} \\
        Infectious Duration (Days) &           5 &      5 &               4    & {4, 5, 6} \\
            Latent Duration (Days) &           4 &      4 &               3    & {3, 4, 5} \\
               Mask Wearing &           0.41 &     0.335 &               0.36 & 
               \makecell[tl]{0.2, 0.31, 0.335,\\0.35, 0.36, 0.385, 0.41} \\
                Remote Work &           0.10 &     0.10 &               0.15 & {0.1, 0.15, 0.2} \\
           Community Distancing &           0.20 &     0.15 &               0.20 & {0.1, 0.15, 0.2} \\
       Quarantine Adherence &           0.60 &     0.60 &               0.70 & {0.60, 0.70} \\
     Base Transmission Rate &           0.20 &     0.175 &               0.125 & \makecell[tl]{0.1, 0.125, 0.15,\\0.175, 0.2} \\
              Immune Escape &           0.40 &     0.40 &               0.80 & {0.4, 0.5, 0.6, 0.7, 0.8} \\
        Antigen Tests Per Day &          \multicolumn{3}{c}{20 million} & {10, 15, 20 million} \\
        Presymptomatic Duration (Days) &          \multicolumn{3}{c}{2} & {1, 2} \\
        School Closure & \multicolumn{3}{c}{0.0} & {0.0, 0.1} \\
\bottomrule
\end{tabular}
\caption{Variation in epidemiological, social, and policy conditions across 3 baseline scenarios. We also include values searched over in the Calibration Search Space column.}
\label{tab:baselines}
\end{table}
Epidemiological models are often characterized by their $R_0$: the measure of expected secondary cases in a completely susceptible population. In an agent-based simulation, calculating $R_0$ does not have a closed form solution. Following standard best practice \parencite[]{germann_mitigation_2006}, we calibrate our $R_0$ value by running many repetitions of TRACE-Omicron with an initially infected ``index agent.'' We  estimate $R_0$ by calculating the average number of infections \parencite{newman_networks_2018} generated by second generation infectives (as the first generation are atypical) \parencite{andersson_epidemics_1997}, specifically we do this calculating the ratio of tertiary to secondary cases. Our chosen base transmission probabilities of 0.2, 0.175, and 0.125 correspond to $R_0$ values of about 8, 7.7, and 6.9 respectively.

\begin{figure}[t]
\centering
\subfigure[]{\includegraphics[width=0.75\textwidth]{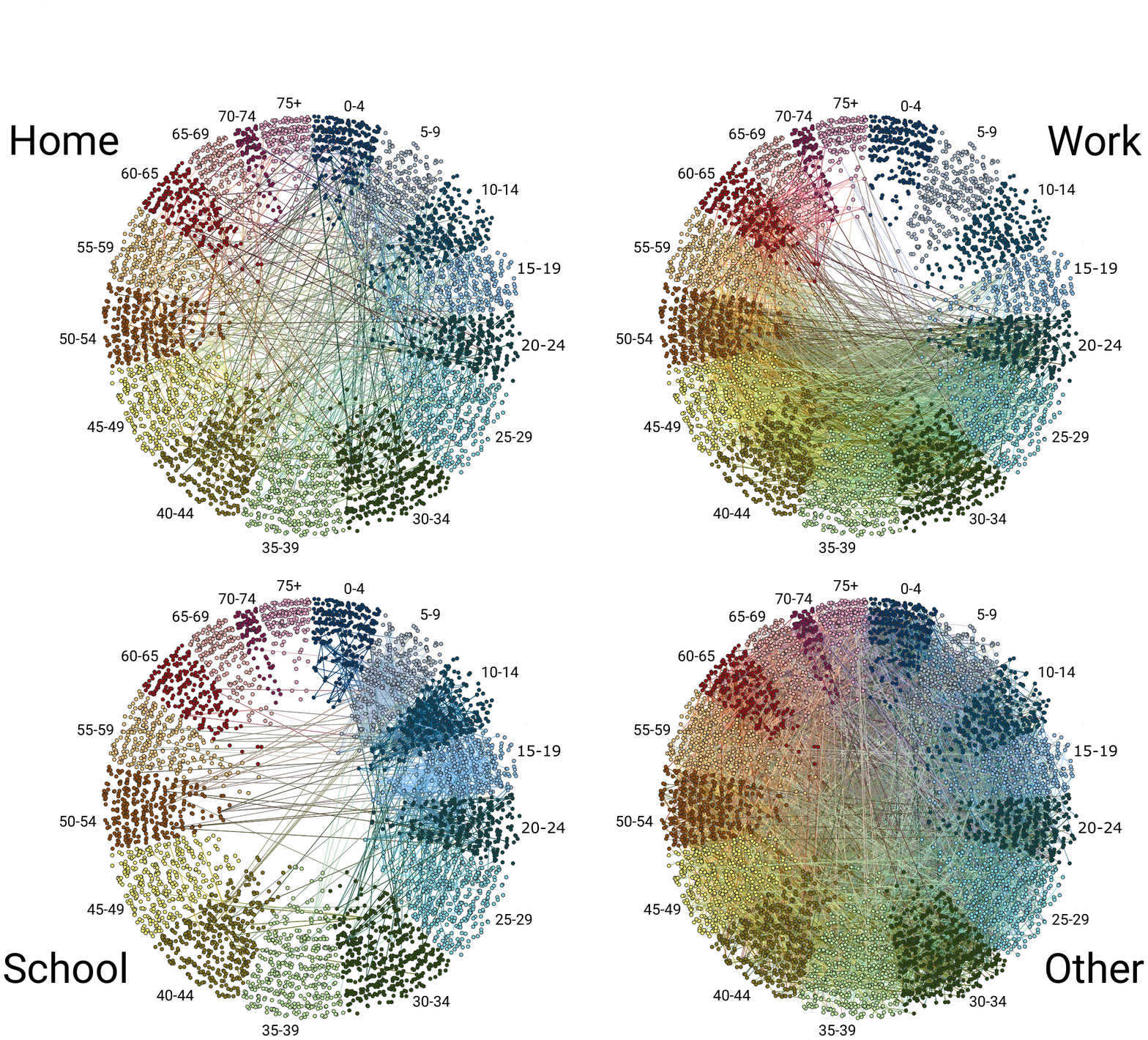}}\\
\subfigure[]{\includegraphics[width=0.55\textwidth]{figures/ego_network.pdf}}
\caption{\textbf{(a)} A visual representation of the contact network for the populations used in simulations. Each subfigure shows the contacts in a given setting for the same sample of 10\% of nodes. Node layout is according to a novel algorithm which bins nodes angularly according to categorical variable (age group in this case) and distance from the center is inversely proportional to degree.  \textbf{(b)} Contacts of an example infectious agent from subfigure a, detailing its variable and the variables of its contacts which affect the probability of an infection. Edge width denotes infection probability if susceptible, text color denotes infection state.
}
\label{fig:networks}
\end{figure}
\clearpage
\subsection{Population Structure}

To balance computational feasibility and observational power, TRACE-Omicron uses a population of 50,000 agents, with social settings (where they live, work, and attend school), demographic attributes, vaccination history, and infection states. Contacts were generated using SynthPops \parencite{noauthor_synthpops_2022}. First widely used in \cite{kerr_covasim_2021} where it is explained in detail, Synthpops is an open-source model for generating realistic synthetic contact networks of specific populations using input data from those populations. After inputting data on nationally representative population age distributions \parencite{noauthor_s0101_nodate}, employment rates stratified by age \parencite{bureau_civilian_nodate}, school enrollment rates by age \parencite{bureau_school_nodate}, household head age distributions by family size \parencite{bureau_families_nodate}, and household, school, and workplace size distributions \parencite{bureau_families_nodate,noauthor_digest_nodate,bureau_number_business_nodate} from federal surveys, Synthpops generates contact structures which are realistically transitive by generating synthetic locations for school, work, and home and assigning contacts based on these locales.  Community contacts are generated at random with quantity drawn from a negative binomial distribution, similar to previous work that also used Synthpops \parencite{kerr_covasim_2021}. These four settings: home, school, work, and community, illustrated in Fig.~\ref{fig:networks}, stratify the contact lists, and agents only have the potential to infect contacts from settings they are not currently disconnected from due to distancing policy (see Modeling Interventions).

\subsection{Infector-Infectee Dynamics}

Besides disease state, agents have a number of other statuses attributed to them which can affect their contact structure and the probability of infection given an interaction. Some agent variables can affect contact structures in ways that impede infection opportunities. These include quarantine, and social distancing measures of working from home, school closure, and not engaging with social community contacts. Other agent variables affect probability of an infection given an interaction between an infectious agent and a susceptible agent. These include masking status (of both the infectious and susceptible agents) as well as vaccine-related attributes (vaccination status, type of vaccine administered, time since vaccine, and number of vaccinations). 

\subsection{Modeling Interventions}

To study ways to mitigate COVID-19 spread, TRACE-Omicron models an array of policy interventions across a high-dimensional parameter space. We simulate policies at baseline, representing the policy context in the United States in late 2021, as well as levels of increasing policy strength to model policy counterfactuals: preventative actions that might have been taken before or during the Omicron wave, but were not. Specifically, we model vaccination and boosting, masking, testing and contact tracing, and social distancing measures. 

Vaccines and boosters are one of the most critical tools for managing COVID-19 \parencite{cdc_vaccines_stay_up_to_date_covid-19_2022}. The protective impact of vaccination and boosting, as interventions, are driven by several key implementation details: who gets vaccinated, when they get vaccinated, what vaccine they receive, how effective that vaccine is at preventing infection, its efficacy against transmission, and how long this protection lasts. Further, vaccine effectiveness and duration of protection is specific to each individual vaccine formulation (e.g., Moderna) and each variant of COVID-19, and much effort has been made to estimate these quantities across variants and vaccine products \parencite{international_vaccine_access_center_ivac_effectiveness_nodate}. 
Using the CDC’s time-series data of vaccination in the United States, vaccines are distributed proportionally by age group among the agents in the simulation \parencite{cdc_vaccination_data}. When agents are vaccinated, they receive two multipliers, one which reduces the probability of transmission, should they experience a breakthrough infection, as well as one reducing the probability of infection relative to an unvaccinated agent. This protection against infection wanes over time, and is updated weekly for each agent, based on real-world vaccine-effectiveness studies \parencite{robles-fontan_effectiveness_2022}. 

Agents can also be boosted 180 days after they have been vaccinated, if their age group is eligible \parencite{scobie_update_2021}. Boosting restores an agent’s vaccine efficacy multipliers to their original values. TRACE-Omicron investigates policy questions regarding how many agents needed to be vaccinated and boosted before the Omicron wave, as well as the estimated efficacy of a faster distribution of available doses. Boosting policies are implemented identically to vaccination policies, but boosting and vaccination are separate interventions in TRACE-Omicron.

A proportion of agents wear masks, which, much like vaccines, provide protection in two directions, each affected by mask efficacy: reducing their probability of infection relative to an unvaccinated agent, and their probability of their contact being infected. There are two masking-focused policy levers: the rate of mask wearing in the population, and the type of masks agents wear. For computational feasibility, agents who wear masks are assumed to wear them at all times. To more closely match mask-wearing in TRACE-Omicron to observational data, we adjust our estimates of baseline mask-wearing downward by the population’s average number of non-home contacts \parencite{institute_for_health_metrics_and_evaluation_ihme_2020,noauthor_synthpops_2022,reinhart_open_2021}. We implement two types of masks in TRACE-Omicron, one which represents standard cloth masks, and one which more closely represents N-95 masks, which have higher protection against both infection and transmission \parencite{koh_outward_2022,ueki_effectiveness_2020}. 

Each day, a sample of non-quarantined agents, subject to test availability, are tested. Agents who test positive are asked to quarantine and report their contacts \parencite{centers_for_disease_control_and_prevention_covid-19_quarantine}. Testing includes polymerase chain reaction (PCR) and antigen (rapid) testing, each with their own false positive and false negative rates \parencite{surkova_false-positive_2020,pray_performance_2021,prince-guerra_evaluation_2021,pecoraro_estimate_2022,cdc_antigen_guidance}. PCR tests are first distributed to a sample of symptomatic agents, and any excess capacity is allocated randomly. Antigen tests are distributed to a random sample of contacts of agents who have tested positive recently, and excess capacity is allocated randomly. A similar procedure is used to simulate contact-tracing, where subject to contact tracing capacity, random contacts of symptomatic agents are asked to quarantine.  As policy interventions, we vary the amount of PCR and antigen tests and contact tracing capacity in the model, and specific policies are represented as multipliers above baseline testing and tracing  levels observed during the Omicron wave \parencite{cdc_covid_data_tracker}.

Social distancing measures encompass individual quarantine or reducing the number of contacts agents have in a given setting. Agents who believe they are infected, receive a positive COVID-19 test, or are contact-traced and asked to self-isolate for a period of time based on whether they are symptomatic. Agents adhere to quarantine directives probabilistically, representing their willingness and ability to do so \parencite{ahmed_paid_2020,tseng_patients_2021,edwards_influenza_2016}. TRACE-Omicron implements business and school closures, and community distancing via removing contacts agents have through the corresponding setting (work, school, or community/other) during the period that closure/distancing is in effect \parencite{bureau_of_labor_statistics_measuring_nodate,national_center_for_education_statistics_2022_nodate,google_covid-19_2020}. A visualization of TRACE-Omicron’s contact structure is shown in Fig.~\ref{fig:networks}.

\subsection{Policy Intervention Intensities}

We generate combinations of our selected policies at multiple levels of intervention. We synthesize these policy combinations into a subset of available policies, and consider singular interventions (i.e., substantial increases in one policy dimension such as testing or social distancing) as well as policy mixtures (i.e., moderate increases in several policy dimensions).  We denote somewhat large increases in policy interventions with a ``+'' and even more intense increases with ``++.'' We summarize the considered policies, with detailed descriptions, in Table~\ref{tab:policies}.

\begin{table}
\centering
\begin{tabular}{llll}
\toprule
\textbf{Policy} & \textbf{Baseline} & \textbf{+} &  \textbf{++} \\
\midrule
\makecell[l]{\textbf{Vaccination} \\ {}  {} Ages 18+ \\ {}  {} Ages 12-17 \\   {}  {} Ages 5-11 \\  {}  {} Ages 0-5} & \makecell[l]{\\0.73 \\0.54 \\0.15 \\ -} & \makecell[l]{\\0.73 \\0.73 \\0.15 \\-\\} & \makecell[l]{\\0.73\\0.73\\0.73\\0.15 \\} \\
\textbf{Boosting} & 0.2 &                                 2x &                                        4x \\
\makecell[tl]{\textbf{Testing}\\{} {} PCR \\ {} {} Antigen} &   \makecell[tl]{\\1.9 million \\20 million} &                             \makecell[tl]{\\2x\\2x}  &  \makecell[tl]{\\5x\\4x}  \\
\makecell[tl]{\textbf{Mask Efficacy} \\ {} {}  Against Infection \\ {} {} Against Transmission} &                                           \makecell[tl]{\\ 40 \\60} &                                           \makecell[tl]{\\ - \\} &                                          \makecell[tl]{\\57 \\ 76} \\
\makecell[l]{\textbf{Mask Wearing}\\\\} & \makecell[tl]{Best Fit: 41\\Tractable Strain: 33.5 \\High Immune Escape: 36} &                                           - &                                   70 \\
\makecell[l]{\textbf{Social Distancing} \\(Work, School, Community) \\} & \makecell[tl]{Best Fit: 0.10, 0, 0.20\\Tractable Strain: 0.10, 0, 0.15\\High Immune Escape: 0.15, 0, 0.20} &                                 0.35, 0.30, 0.30 &                                                  0.60, 0.60, 0.50 \\
\bottomrule
\end{tabular}
\caption{Summary of Policy interventions and intensities. The ``+'' column represents a somewhat large increase in the strength of the policy relative to baseline, and ``++'' represents a much larger increase. Vaccination, Boosting, and Mask Wearing represent the proportion of the population receiving that policy. Testing represents the amount of tests available per day, by test type. Mask Efficacy represents the percentage reduction in infection probability in an interaction. Social Distancing represents the percentage reduction in contacts for each social setting. Across our three calibration scenarios, we estimate slightly different baseline values for Mask Wearing and Social Distancing, which are displayed here. The parameters displayed here may be combined with Table S2 for a full characterization of the model parameters.}
\label{tab:policies}
\end{table}
\subsection{Epidemiological Counterfactuals}

Finally, we explore the results of policy interventions to study the Omicron wave if a different variant of SARS-CoV-2 had become dominant, for example, one that could evade immunity more effectively than Omicron or was even more contagious. In this way, we can gain insight into whether or how policy impacts might differ against a future variant of the virus. 

In these runs we chose base transmission rates 0.09, 0.2 (the calibrated baseline value from the Best Fit calibration), and 0.55, which correspond to $R_0$ values of 6, 8, and 10. 

\subsection{Computational Implementation}

TRACE-Omicron was programmed in Python 3.9. We conducted 88,128,000 model runs exploring 46,080 parameter combinations. Computations were performed, in part, on the Vermont Advanced Computing Core and the Washington University Scientific Compute platform. Source code of TRACE-Omicron is available on a public repository.

\section{Results}

\subsection{Retrospective Analysis}

\begin{figure}[htbp]
    \centering
    \includegraphics[width=0.8\textwidth]{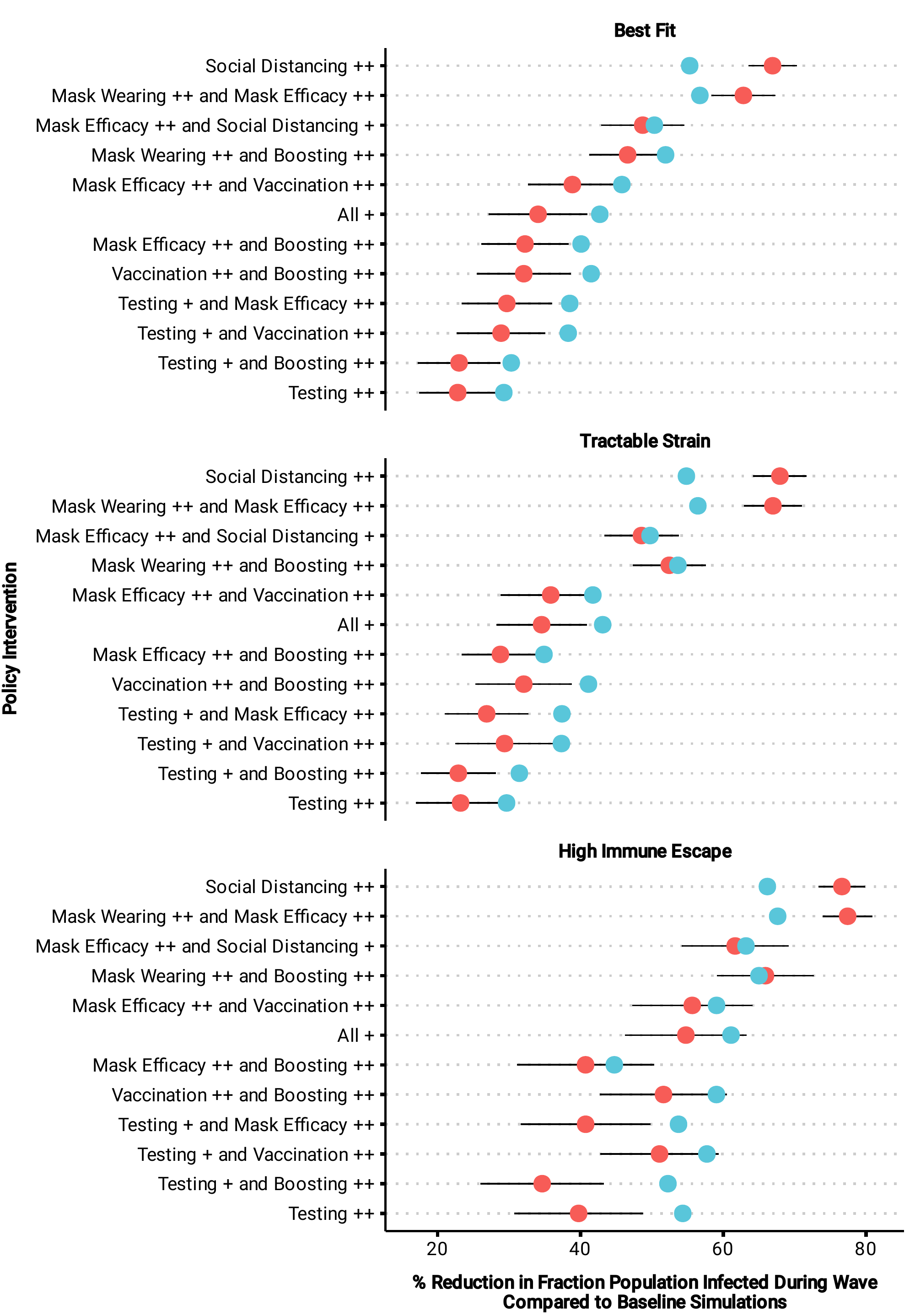}
    \caption{Summary of selected policy interventions across baseline scenarios. Red dots represent the average percentage reduction from baseline in cumulative infections (with 
95\% confidence intervals), and blue dots represent the maximum number of infected agents at one time across model repetitions.}
    \label{fig:pointline}
\end{figure}

Given that TRACE-Omicron achieves generative sufficiency in recreating the number of infections during the Omicron wave, we analyze what policies may have mitigated the spread of COVID-19 if implemented prior to or during the Omicron wave of late 2021 and early 2022. A sample of policy interventions are summarized in Fig.~\ref{fig:pointline}. Our analysis shows multiple policy options allowing for significantly lower rates of cumulative infection and peak surge levels during the Omicron wave than baseline. In particular, policies such as Social Distancing ++, Mask Wearing ++ and Mask Efficacy ++, Mask Efficacy ++ and Social Distancing +, and Mask Wearing ++ and Boosting ++, all reduced simulated cumulative infections by 40\% or more. We also observe combinations of less disruptive policies that project similar reductions in infection to mass social isolation policies, which are effective but socially disruptive. 

\begin{figure}
    \centering
    \includegraphics[width=\linewidth]{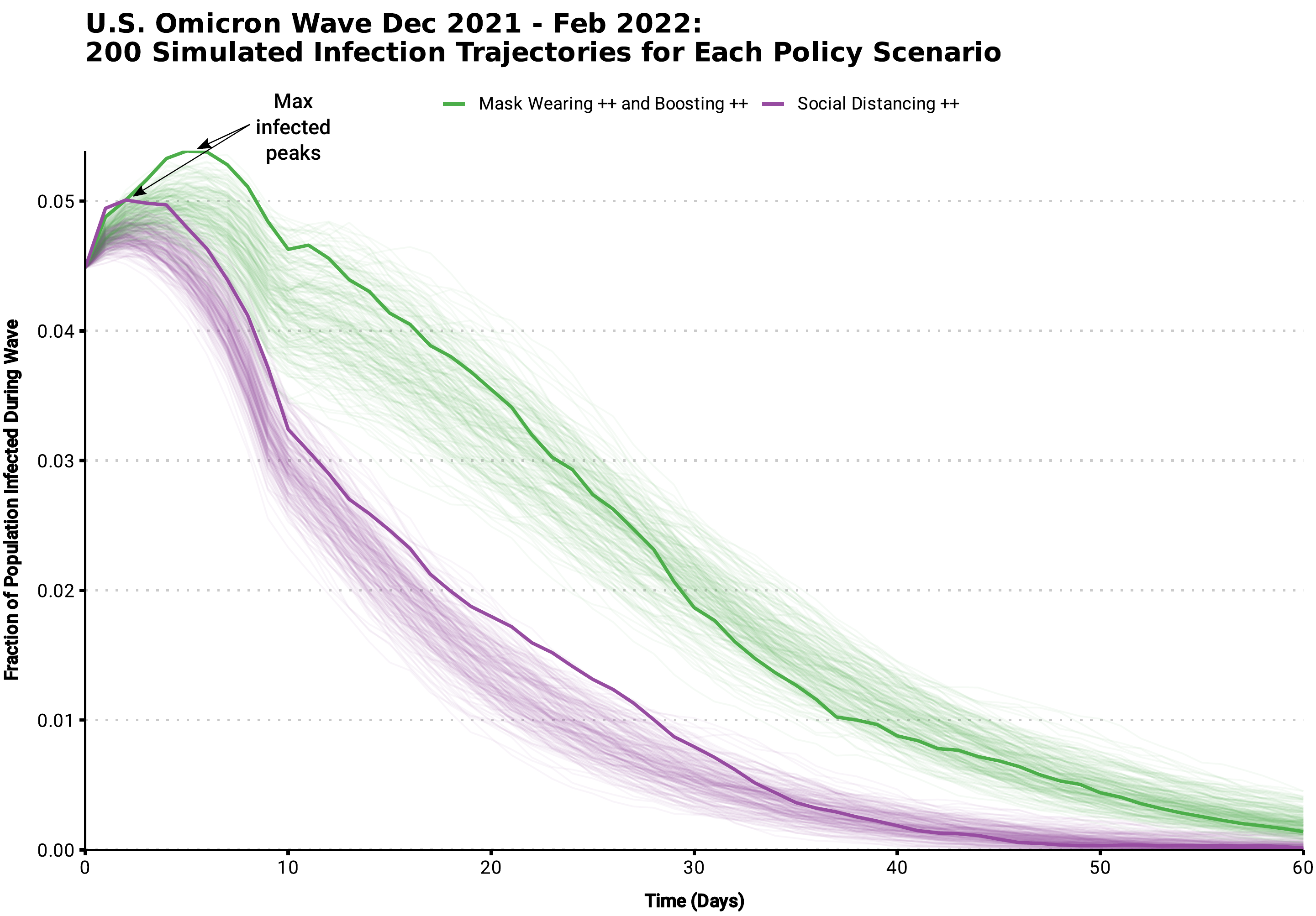}
    \caption{Comparison of Boosting ++ and Mask Efficacy ++ vs. Social Distancing ++ under Best Fit calibration. The highlighted curves represent the model runs resulting in the maximum number of infected agents at one time.}
    \label{fig:showpeaks}
\end{figure}

Selecting optimal policies depends not only on acceptable cost but on the choice of metric to evaluate the impact of these policies. The total size of the wave as a number of cases and the height of the epidemic peak are not fully correlated. While the ultimate goal may be to reduce the total number of cases, doing so without exceeding a threshold of concurrent cases that overwhelms the healthcare system is often a priority for policymakers. Notably, the combination of Mask Wearing ++ and Mask Efficacy ++ policies are a close second to strong social distancing in terms of limiting the total number of cases, yet the combinations of these strong masking procedures better reduce the peak of the wave compared to social distancing alone. They might therefore be preferred to lighten the load on the healthcare system. Conversely, a combination of Mask Wearing ++ and Boosting ++ achieves a similar epidemic peak as strong social distancing alone, but leads to a longer wave with more total cases (Fig. ~\ref{fig:showpeaks}).

\subsection{Synergistic Effects}

Our results also show the potential for synergistic policy effects: combinations of increased use of high-quality (e.g., N95) masks and increased testing were estimated to be about as effective and increased vaccination and booster uptake. Vaccinating children under 18 at the rate of adults would have also been estimated to largely decrease disease spread. Both boosting and vaccination policies were even more effective when combined with other policies, such as masking or testing.

\begin{figure}
    \centering
    \includegraphics[width=\linewidth]{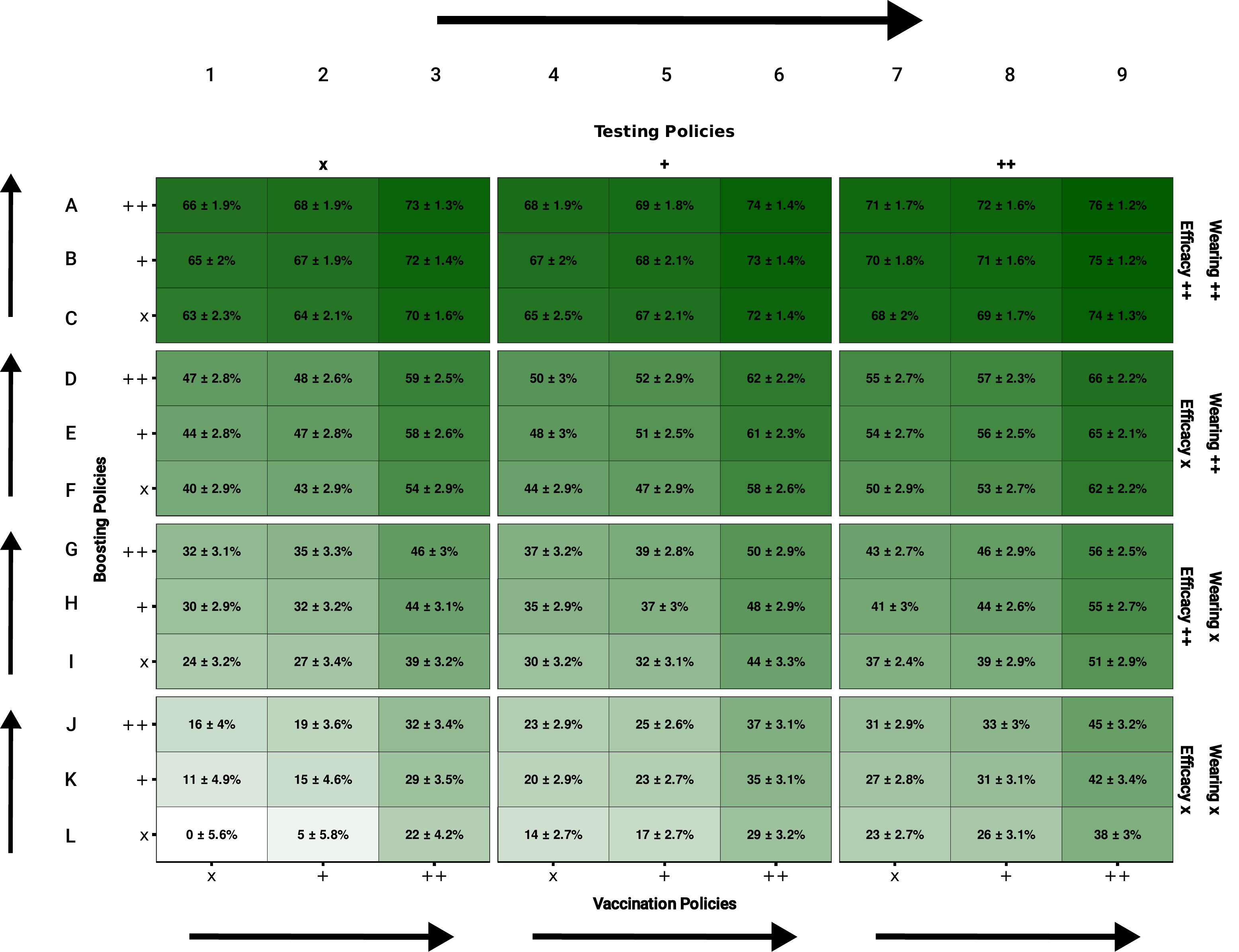}
    \caption{Summary of policy intervention landscape under the Best Fit baseline. Cells represent the average percentage reduction from baseline in cumulative infections (with 95\% confidence intervals). All simulations shown hold social distancing policies (remote work, school closure, and community distancing) to their baseline levels. Magnitudes of policy interventions are described in Table~\ref{tab:policies}.}
    \label{fig:heatmap-best}
\end{figure}

To further explore policy synergies, combinations of vaccination, boosting, testing, and masking policy strategies for our Best Fit are displayed in Fig.~\ref{fig:heatmap-best}. Additional heatmaps are available in the Supporting Information. In each heatmap, every 3x3 matrix (sub block) represents a specific combination of social distancing policy (top) and masking policy (right hand side). In each of these sub blocks, cells represent a given combination of vaccination policy (bottom) and boosting policy (left hand side).

Overall, the Best Fit results shown in the heatmap suggest that it is possible to achieve large-scale reductions in COVID-19 infections through multiple means. This includes policies with no increases in social distancing or quarantine adherence, even without increasing policies to the highest levels we simulated in TRACE-Omicron. More specifically, 49 of 108 available policy interventions would have decreased cumulative infections by 50\% of infections observed at baseline.

Mixtures of several policies can substitute, or even outperform large increases in individual policies. Individually, Boosting ++, Vaccination ++, and Testing ++ on their own lead to cumulative infection reductions from baseline of 16\% (cell J1), 22\% (cell L3) and 23\% (cell L7), respectively. A mixture of all 3 policies at lower intensities (Boosting +, Vaccination +, Tests +) resulted in a simulated cumulative infection reduction of 23\% (cell K5), as good or better than any of the larger intervention scenarios alone. Supplementing this policy mixture with Mask Wearing ++ or Mask Efficacy ++ resulted in simulated cumulative infection rates of 51\% (cell E5) and 37\% (cell H5), respectively. Overall, notice how in each 3x3 matrix, the middle cell representing a mixture of policy interventions is better than the top left corner representing doing more of one policy and not increasing others. Mixtures of policies outperform policies focused on one intervention specifically.

\subsection{Specific Effects: Masking and Timing}

\begin{figure}
    \centering
    \includegraphics[width=\linewidth]{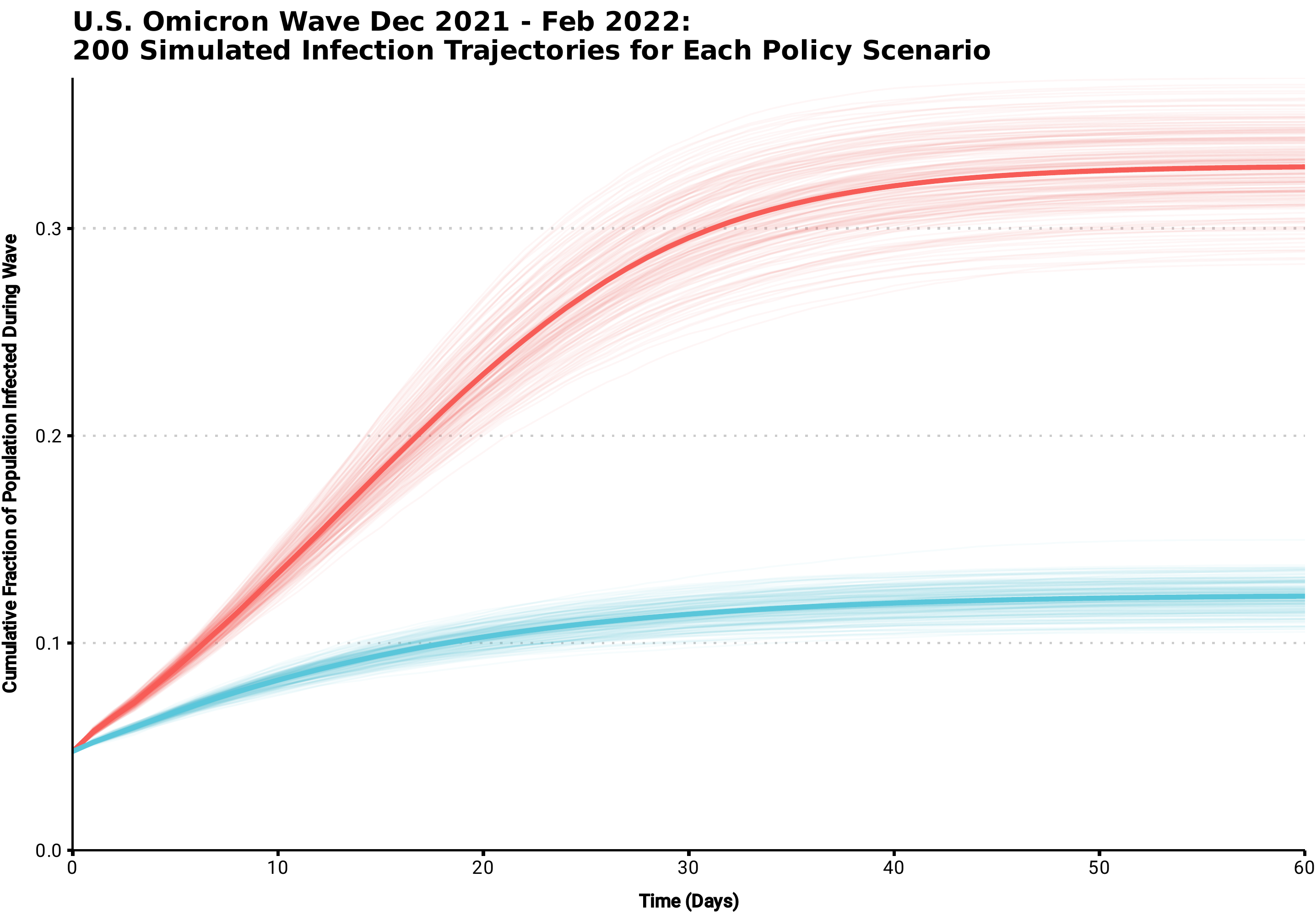}
    \caption{Comparison of a mask-focused intervention (Mask-Wearing ++ and Mask Efficacy ++) to baseline under Best Fit calibration.}
    \label{fig:Baseline vs. Mask Efficacy and Mask Wearing ++}
\end{figure}

Digging into specific policies, we find that masks, especially high-quality ones, remain powerful tools in mitigating disease spread. Simulated cumulative infection rates under all Mask Wearing ++ and Mask Efficacy ++ scenarios were reduced on average by 63\% (cell C1) representing a substantial reduction from baseline. We also present the results of a time-series comparison of a mask-focused policy vs baseline in Fig.~\ref{fig:Baseline vs. Mask Efficacy and Mask Wearing ++}. We see that the number of simulated infections, the rate at which agents are infected, and the variance of model simulation outcomes, are all much lower under a policy intervention with high proportions of effective masking.

\begin{figure}
    \centering
    \includegraphics[width=\linewidth]{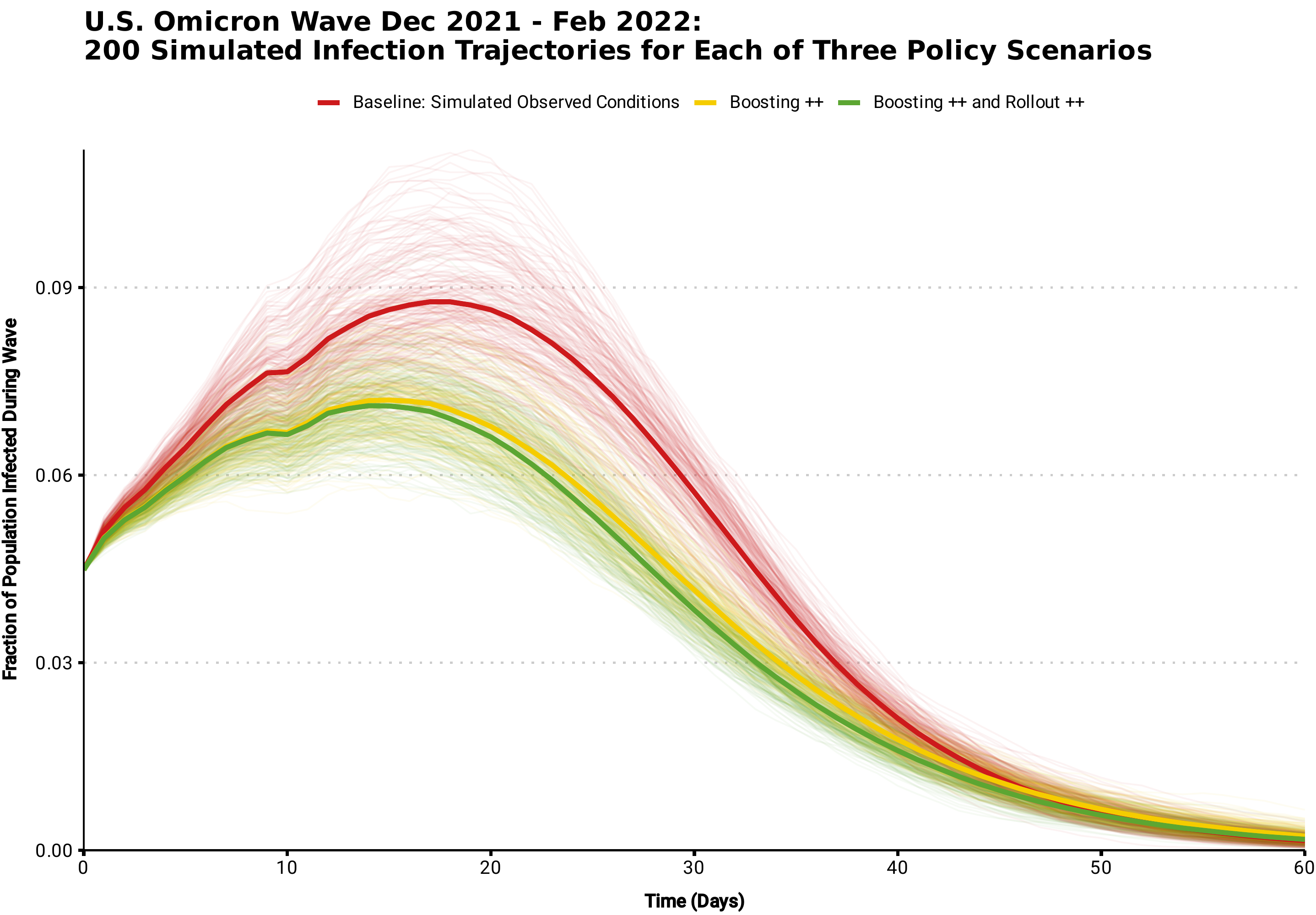}
    \caption{Comparison of baseline, Vaccination ++ and Boosting ++, Vaccination ++ and Boosting ++ and Rollout ++ under Best Fit calibration.}
    \label{fig:High Boosting and Vaccine Rollout}
\end{figure}

The timing of policies is also crucial for pandemic response. In Fig.~\ref{fig:High Boosting and Vaccine Rollout}, we investigate the impact having a higher boosted population prior to the start of the Omicron wave, along with an additional comparison line which adds the marginal effect of adding faster booster and vaccine distribution after the start of the Omicron wave. We find that pandemic preparedness (i.e., the proportion boosted before an outbreak) to be the more important factor in reducing simulated cumulative infections compared to ramping up boosting and vaccination efforts after disease spread has already begun. 

\subsection{Robustness Under Alternate Baselines and Epidemiological Counterfactuals}

The results highlighted above are generally robust under a suite of different calibration baselines, but there are key differences worth highlighting in the High Immune Escape scenario. Mainly, with High Immune Escape, we find that most policies are much more effective, especially Social Distancing ++ as well as the combination of Mask Wearing ++ and Mask Efficacy ++. This result can be explained by the fact that if the Omicron wave featured a high immune escape rate, then its susceptible pool would have been effectively much larger than expected and its effective transmission rate lower and easier to control. Following the same logic, we actually see an increase in the importance of vaccination under a high immune escape rate. This might appear paradoxical but it is again due to the reduction in natural learned immunity in the population. As a caveat, it is also worth noting that all scenarios under the High Immune Escape scenario produce larger confidence intervals, further confounding our ability to produce a simple ranking of policy combinations.

Despite the remaining uncertainty in the epidemiological conditions of the early Omicron variants, our comparison across the alternate baselines suggests a consistent encouraging message that given proper preparation, variants like Omicron can be mitigated. However, our concerns extend beyond Omicron-like variants. In addition to our policy counterfactuals, we also analyzed epidemiological counterfactuals, simulating the same policy interventions under scenarios alternate to the observed Omicron wave in immune evasion and transmissibility.

\begin{figure}
    \centering
    \includegraphics[width=\linewidth]{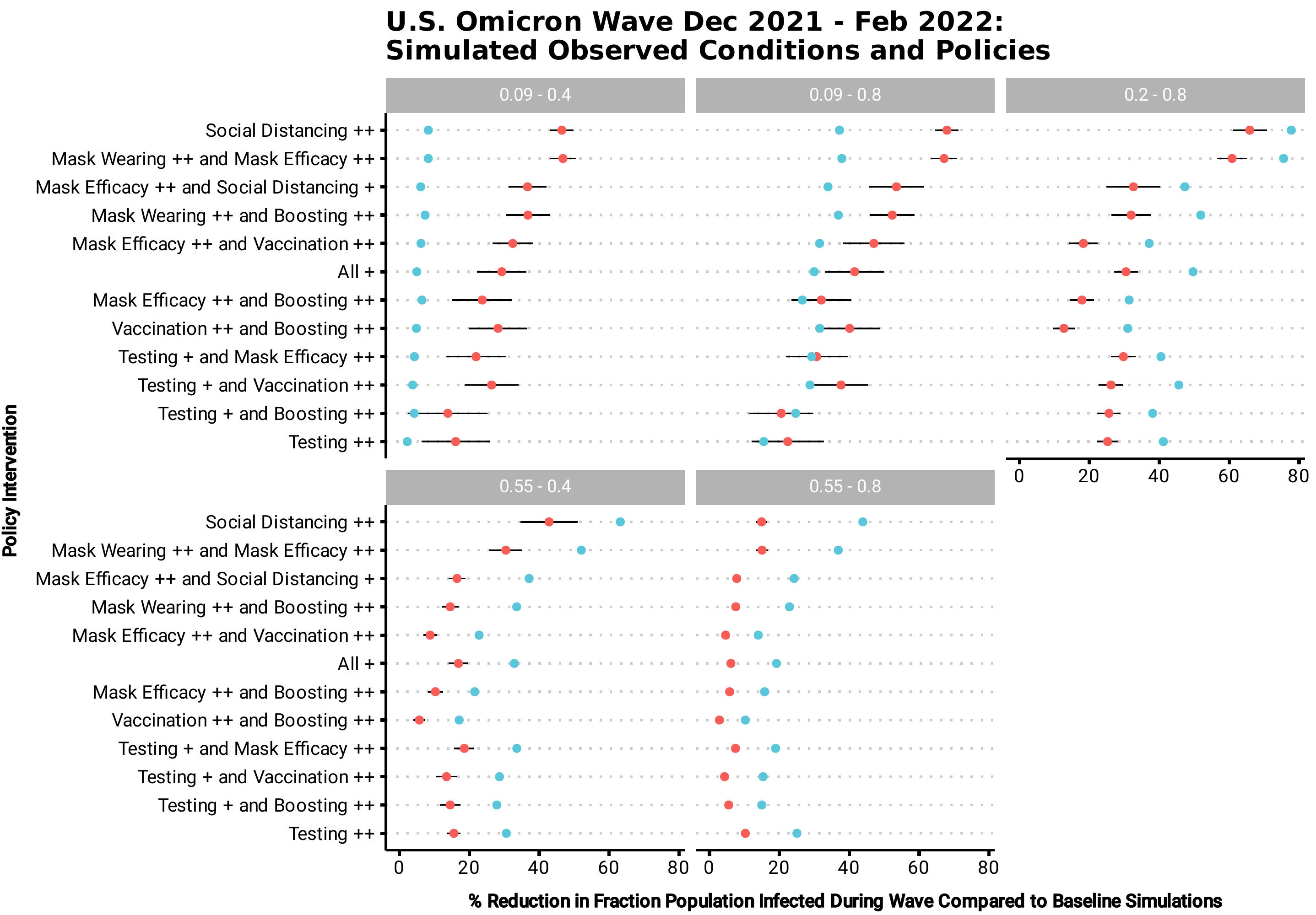}
    \caption{Summary of selected policy interventions by epidemiological scenario (base transmission and immune escape rates) using Best Fit as baseline. Red dots represent the average percentage reduction from baseline in cumulative infections (with 95\% confidence intervals), and blue dots represent the maximum number of infected agents at one time across model repetitions.}
    \label{fig:PointandLinePlotAllEpi}
\end{figure}

Our key findings (Fig.~\ref{fig:PointandLinePlotAllEpi}) are that while the overall effectiveness of policy interventions changes under different epidemiological scenarios, the rank-ordering of these policies is largely unchanged. In particular, increased levels of effective mask-wearing are nearly as effective at reducing cumulative infection as large increases in social distancing. These interventions are also much more effective at reducing the maximum number of infected agents at one time, especially for more pessimistic epidemiological scenarios (with high virus transmission and immune escape). Under the two most pessimistic scenarios, we observe that Vaccination and Boosting focused strategies alone are the least effective among the selected policy interventions at lowering cumulative infections. This is due in part to lower baseline levels of vaccination and boosting in the United States, as well as the waning effectiveness of vaccines against infection.

However, small increases in all policies (All +) can still outperform large interventions focused on one of two policies. This is especially apparent in the scenario with a more transmissible variant with the same estimated levels of immune escape, where the All + scenario is the 4th most effective policy, in contrast to being the 6th most effective at baseline.

Further analyses of the epidemiological counterfactuals are available in the Supporting Information.

\subsection{Summary}

Altogether, we find that ranking of potential policies is a complex task. First, the expected effectiveness of a policy will depend on our current knowledge on the incoming wave, as shown by our epidemiological counterfactuals, especially in those parameterizations with high immune escape. Second, we found that the effectiveness of a policy depends on the synergies between the numerous possible pairings of interventions. Third, the rankings also depend on the preferred metrics of intervention outcomes, or combinations thereof. We here analyzed our results based on multiple combinations of interventions and two key metrics, total cases and epidemic peaks. Despite these complexities, TRACE-Omicron proved a useful tool for generating actionable insights, and we identified several consistent themes across all of the scenarios explored. We find that testing-based strategies alone will likely not be sufficient to contain disease spread, vaccination-based policies are most effective when implemented ahead of time rather than in response to disease spread, and mask-based policies, which can be deployed quickly, tend to be highly effective. Underlying all these analyses, we also see the effectiveness of mixtures of policy interventions, which often outperform aggressive individual policy increases.  We make our complete database of simulation runs available in the hope of encouraging further analysis of the complex interplay between policy combinations and epidemic dynamics.

\section{Discussion}

TRACE-Omicron was designed and analyzed to estimate the effects of policies had they been implemented in late 2021 or early 2022. Given that the emergence of the Omicron variant produced a severe wave of cases in a vaccinated population, studying the impact of putative response strategies that might have been deployed can provide useful insights in advance of future waves. As of early Fall 2022, less than 15\% of Americans have received the newest booster \parencite{cdc_vaccination_data}. Mask wearing is less common than it was before the Omicron wave \parencite{wu_its_2022}. On the positive side, the Omicron wave resulted in a large increase in the frequency of use of at-home tests \parencite{rader_use_2022}.

Our results provide an important path forward in considering the implications of these behavior shifts and policy possibilities for protecting against future waves. We outline, promisingly, how a number of different mixtures of policies can be quite effective – even outperforming large interventions focused on one policy. One such policy that particularly underperforms when used alone is vaccination. Even a massive speedup to the (already historically fast) vaccination pipeline that would have been necessary prior to Omicron to be able to vaccinate young children, and to vaccinate older children and teens at the rate of young adults, still only reduced outbreaks in our model by less than a quarter in the Best Fit and Tractable Strain scenario and while better, perform near the worst of any solo extreme strategy in the High Immune Escape scenario. However, in combination with or in place of vaccines, there are several combinations of alternate strategy suites which yield lower outbreak sizes. This is fortunate in the wake of recent concerns over rapidly evolving variants \parencite{groenheit_rapid_2023} outpacing vaccine advancements as well as vaccine uptake \parencite{lu_covid-19_2022}.

That is not to say however, that vaccination is not critical for controlling this disease, as studies show that vaccinations significantly reduce serious complications and deaths \parencite{arbel_bnt162b2_2021}. Current discussions in this work were focused on the number of cases during the Omicron wave, both in total and at its peak. It is thus an important limitation that we do not directly model severe disease and mortality which are increasingly becoming the key indicator of policies. Even then, the number of cases remains the key output as other metrics like mortality could be extracted by using an appropriate stratified risk of complications. However, some important outcomes such as long COVID or severe post-COVID conditions are still too poorly understood to be modeled directly in this fashion. And current science shows that risks of such complications increase with repeated infections \parencite{al-aly_long_2022}. Without accurate models for individual risk of complications, we will continue to report cases as the main output of TRACE-Omicron.

Like all epidemiological models, TRACE-Omicron utilizes necessary simplifications, and cannot fully express all aspects of human behavior and interaction. Specifically, TRACE-Omicron does not model social influence on policy adoption. For example, it is well-known that mask-wearing in the United States has been a divisive issue, and differs across social and demographic groups \parencite{mallinas_what_2021}. We also do not model treatment of COVID-19, specifically therapeutics under Emergency Use Authorization in the United States such as Paxlovid, since these were not widely available during the Omicron wave and were only recommended for a subset of the population \parencite{department_of_health_and_human_services_covid-19_therapeutics}. Finally, we do not model competing variants or viral evolution in TRACE-Omicron.

The range of potentially equally effective mixed strategies which avoid socially disruptive lockdowns is promising from an economic and policymaking standpoint. We identify sets of direct substitutes with quantitative equivalency in epidemiological containment; these could be analyzed further from the perspective of economic or social cost, equity concerns, or policy constraints.  To further this goal, we provide, as open data, the complete set of 88,128,000 model runs with daily agent state counts, including 46,080 policy combinations (a subset of which are covered here) that can be used for systematic analyses of economic or social cost (see Data Availability).  Additionally, the computational laboratory our open-source code provides can be used for analyses of more specific or more extreme parameterizations not included in our dataset.

TRACE-Omicron also offers substantial scope as an extensible framework for future research, due to the agent-based, object-oriented, modular nature of the contagion dynamics, population, and policy representation. Additional dynamics can be added to study multiple contagion systems like the current epidemics of COVID-19, influenza and Respiratory Syncytial Virus (RSV) \parencite{tanne_us_2022}, or multiple competing variant systems. Given accurate data, any number of demographics can be included in our representation of the population, as well as correlations between demographics and behavior (e.g. compliance) or risk (e.g. RSV in young children). Additional policies can be simulated. This includes policies which could take advantage of the site-based framework of the population (e.g. ventilation system infrastructure) \parencite{risbeck_modeling_2021}, as well as policies which take advantage of the network aspect of the contact structure such as network-informed targeted immunization strategies \parencite{rosenblatt_immunization_2020,pastor-satorras_immunization_2002}.

The individual-level granularity of our ABM design, coupled with additional demographic input data, presents an opportunity to take health equity into consideration when weighing future response strategies. The negative impact of COVID-19 in the United States has been greatest for those from racial and ethnic minority groups \parencite{romano_trends_2021}. We provide a powerful tool for policymakers who wish to prevent those inequitable patterns being repeated during future pandemic waves. For example, although social distancing might be protective overall for the population, future model application can be used to determine whether and to what extent exposure risks borne by ``frontline'' workers who cannot work remotely \parencite{do_us_2021} might nonetheless drive disturbing racial disparities in harm.

\medskip
\textbf{Supporting Information} \par 
Supporting Information is available from the author awaiting permanent release.

\medskip
\textbf{Data Availability} \par 
Data are available upon request awaiting permanent release, whereupon they will be housed in an open repository.

% Acknowledgements
\medskip
\textbf{Acknowledgements} \par %delete if not applicable))
The authors thank Joe Ornstein, one of the original developers of TRACE, Jane Lydia Adams for consultation about data visualization, and Guillaume St-Onge for consultation about calculations regarding epidemics on networks. Core funding for the TRACE project was provided by The Brookings Institution Economic Studies Program. S.F.R. is supported as a Fellow of the National Science Foundation under NRT award DGE-1735316. L.H.-D. acknowledges support the National Institutes of Health 1P20 GM125498-01 Centers of Biomedical Research Excellence Award. The authors acknowledge the Vermont Advanced Computing Core at the University of Vermont for providing High Performance Computing resources. The authors acknowledge the Research Infrastructure Services (RIS) group at Washington University in St. Louis for providing computational resources and services needed to generate the research results delivered within this paper. URL: \url{https://ris.wustl.edu}

% References
\medskip

% Use the following code if you wish to generate your bibliography with BibTeX;
% replace the string ''MSP-template'' below with the name(s) of
% the BibTeX data base(s) you want to use.
% The resulting bibliography-output (the content of the .bbl file)
% must be pasted back into this file before submission.
% Please also include your BibTeX data base file(s) in your submission
% so that we can re-run BibTeX if necessary.
%
\newpage
\printbibliography
\end{document}